\documentclass[10pt, twocolumn, draft, showpacs, amsmath,amsfonts, amssymb,eqsecnum, prd]{revtex4}

\usepackage[english]{babel}

\def\a{\alpha}
\def\b{\beta}
\def\g{\gamma}
\def\d{\delta}
\def\ep{\varepsilon}

\def\a{\alpha}
\def\b{\beta}

\def\g{\gamma}
\def\d{\delta}
\def\g{\gamma}

\def\J{{\mathcal{J}}}

\def\F{{\cal{F}}}
 \def\H{{\mathcal{H}}}

  \def\A{{\mathcal{A}}}

\begin{document}
 \title{Dispersion relation for anisotropic media}
 \author{Yakov Itin}

\affiliation{Institute of Mathematics, The Hebrew University of
  Jerusalem \\ and Jerusalem College of Technology, Jerusalem,
  Israel. \\ email: {\tt itin@math.huji.ac.il}}

\pagestyle{myheadings}
\markboth{Yakov Itin} {Yakov Itin \qquad
Dispersion relation for anisotropic media}


 \begin{abstract}
The electromagnetic wave propagation  in an anisotropic dielectric media with two generic matrices $\ep^{ij}$ and $\mu^{ij}$ of permittivity and permeability is studied.  
These matrices are not required  to be  symmetric, positive definite, and even invertible. In the framework of a metric-free electrodynamics approach, compact tensorial dispersion  relation  is derived. The resulted formula are useful for a theoretical study of electromagnetic wave propagation in a classical media and in a modern type of media with a generic linear constitutive relation including metamaterials. 
\end{abstract}
\pacs{03.50.De, 46.05.+b, 14.80.Mz}

\date{\today}
\maketitle

\section{Introduction }
Development of the modern microscopic technology (nano-technology) provides a possibility to manufacture materials of a rather non-ordinary electromagnetic parameters. This situation recall the theoretical  investigation of wave propagation in a media  with a generic constitutive law. 

A wide class of media is characterized by a linear constitutive law and in general can be described by four $3\times 3$ matrices. Two of these matrices, $\ep^{ij}$ and $\mu^{ij}$ (permittivity and permeability matrices), describe the pure electric and the magnetic properties of the matter. Two additional matrices describe relativity smaller  electric-magnetic cross-term effects. 

The ordinary textbook's description of a media with two anisotropic matrices  $\ep^{ij}$ and $\mu^{ij}$ is based on a diagonalization of one of them \cite{Zom}, \cite{LL}, \cite{Jack}. This algebraic procedure is always possible for a symmetric matrix. Moreover, if both matrices are symmetric and one of them is positive definite, both of them can be diagonalized. Even with this simplification, the corresponded dispersion relation is given in a rather complicated form. Moreover, it is clear that the diagonalization technique  is not applicable in a general case when both matrices  $\ep^{ij}$ and $\mu^{ij}$ are not symmetric nor positive definite. 

In the current paper, we study the wave propagation in a generic media in the framework of premetric electrodynamics approach \cite{Post}, \cite{birkbook}, \cite{Itin:2009aa}. Our final result is a compact form of the dispersion relation. For two generic matrices  $\ep^{ij}$ and $\mu^{ij}$ , it is given by the expression 
 \begin{equation}\label{fin}
 w^4-2\big(\psi^{ij}k_ik_j\big)w^2 +\frac{\ep^{ij}k_ik_j}{det\, \ep}\,\frac{\mu^{mn}k_mk_n}{det\, \mu}=0\,,
\end{equation}
where
\begin{equation}\label{fin0}
\psi^{ij}=\frac 12\epsilon^{imn}\epsilon^{jpq}\ep^{-1}_{nq}\mu^{-1}_{mp}\,.
\end{equation}
The organization of the paper is as follows: In the next section, the metric-free electrodynamics notations is recalled. In Section 3, the covariant metric-free form of the dispersion relation is represented. The main results are given  in Section 4 where several compact forms of the generic dispersion relation are derived. In Section 5, the examples for isotropic, diagonal anisotropic, and non-diagonal (magnetized ferrite) media are represented. 
 
\section{Anisotropic media in the metric-free description}
Let us start with a metric-free four dimensional system of Maxwell equations
\begin{equation}\label{gen-1}
\ep^{\a\b\g\d}\F_{\b\g,\d}=0\,, \qquad \H^{\a\b}{}_{,\b}=4\pi \J^\a\,.
\end{equation}
It includes two antisymmetric tensors --- the {\it field strength} $\F_{\a\b}$ and the {\it field excitation} $\H^{\a\b}$. The Greek indices change in the range $\a,\b,\cdots=0,1,2,3$, the comma denotes the partial derivatives relative to the coordinates $\{x^0,x^1,x^2,x^3\}=\{ct,x,y,z\}$. In sequel, the Roman indices will be used for the spatial coordinates, $i,j,\cdots=1,2,3$.

The $(1+3)$-decomposition of the field tensors reads 
  \begin{eqnarray}\label{gen-2}
  E_i=\F_{0i}\,,\qquad&& B^i=-\frac 12 \ep^{ijk}\F_{jk}\,,\\
  \label{gen-3}
  D^i=\H^{0i}\,,\qquad&&H_i=\frac 12 \ep_{ijk}\H^{jk}\,.
  \end{eqnarray}
The  electric current is given by 
\begin{equation}\label{gen-4}
\J^0=\rho\,,\qquad \J^i=\frac 1c j^i\,.
\end{equation}
In this notation, the system (\ref{gen-1}) is rewritten in the ordinary three dimensional form of Maxwell equations
  \begin{eqnarray}\label{gen-5}
  div\,{\mathbf B}=0\,,\qquad&& curl\,{\mathbf E} +\frac 1c \frac{\partial{\mathbf B}}{\partial t}=0\,,\\
  \label{gen-6}
 div\,{\mathbf D}=0\,,\qquad&& curl\,{\mathbf H} -\frac 1c \frac{\partial{\mathbf D}}{\partial t}=\frac {4\pi}c{\mathbf j}\,.
  \end{eqnarray}

For a dielectric media, two antisymmetric tensor fields are assumed to be linearly related one to another
 \begin{equation}\label{gen-7}
\H^{\a\b}=\frac 12 \chi^{\a\b\g\d}\F_{\g\d}\,.
\end{equation}
The {\it constitutive tensor} $\chi^{\a\b\g\d}$ is antisymmetric in two pairs of indices, so it  has, in general, 36 independent components. 
Such generic constitutive tensor can be represented by four 3-dimensional matrices of 9 independent components. We will use a representation of the form
   \begin{equation}\label{gen-8}
 \chi^{\a\b\g\d}=\left( \begin{array}{cc}
\ep^{ij} & \g^i{}_j \\
\tilde{\g}^i{}_j&\pi_{ij} 
 \end{array} \right)\,.
\end{equation}
In this paper, we restrict to an electromagnetic media which describes by two tensors $\ep^{ij}$ and $\pi_{ij}$. Two additional tensors $\g^i{}_j$ and $\tilde{\g}^i{}_j$ represent the electric-magnetic cross-terms, which are relativity small for most types of the dielectric materials.  We will consider, however, a some type of a {\it generalized anisotropic media.} In particular, we will not require the matrices $\ep^{ij}$ and $\pi_{ij}$ to be  symmetric, positive definite, nor even invertible. Consequently we will use a constitutive tensor of 18 independent components
   \begin{equation}\label{gen-9}
   \chi^{0i0j}=\ep^{ij}\,,\qquad \chi^{ijkl}=-\epsilon^{ijm}\epsilon^{kln}\pi_{mn}\,.
   \end{equation}
   In three-dimensional form, the corresponded constitutive relation is given by 
      \begin{equation}\label{gen-10}
      D^i=\ep^{ij}E_j\,, \qquad H_i=\pi_{ij}B^i\,.
    \end{equation}
   For a regular matrix $\pi_{ij}$, an inverse permeability matrix 
\begin{equation}\label{gen-10x}
\left(\pi^{-1}\right)^{ij}=\mu^{ij}
\end{equation} 
is defined. With this notation, the constitutive relation takes the ordinary form 
   \begin{equation}\label{gen-11}
      D^i=\ep^{ij}E_j\,, \qquad B^i=\mu^{ij}H_j\,.
    \end{equation}
\section{A general dispersion relation}
A covariant dispersion relation for a generic constitutive tensor $\chi^{\a\b\g\d}$ recently accept a considerable interest \cite{birkbook},\cite{Obukhov:2000nw}. Here we briefly recall the necessary notations and the main stages of the derivation as it given in \cite{Itin:2009aa}. 

Our aim is to establish the necessary conditions for existence of  physically non-trivial solutions of the source-free system
  \begin{equation}\label{disp-1}
\epsilon^{\a\b\g\d}\F_{\b\g,\d}=0\,, \qquad \chi^{\a\b\g\d}\F_{\b\g,\d}=0\,.
    \end{equation}
Here the ordinary condition of the geometric optics approximation is accepted. In particular, we consider the media parameters encoded in 
$\chi^{\a\b\g\d}$ as varied slowly relative to the change of the electromagnetic field.  

The first equation of (\ref{disp-1}) has a standard solution in  term of the vector potential $\A_\a$
 \begin{equation}\label{disp-2}
 \F_{\a\b}=\frac 12 \left(\A_{\a,\b}-\A_{\b,\a}\right)\,.
     \end{equation}
Consequently, the second equation of (\ref{disp-1})  takes the form
     \begin{equation}\label{disp-3}
 \chi^{\a\b\g\d}\A_{\g,\b\d}=0\,.
     \end{equation}
Let us look for a solution of this equation in the form of a monochromatic wave ansatz 
  \begin{equation}\label{disp-4}
\A_{\a}=a_\a e^{iq_\b x^\b}\,.
     \end{equation}
We substitute this ansatz into (\ref{disp-3}) and treat the amplitude of the field $a_\a$ and the wave covector $q_\b$ as slow functions of a spacetime point. 
 Consequently, we come to an algebraic system 
  \begin{equation}\label{disp-5}
M^{\a\d}a_\d=0\,
     \end{equation}
with a characteristic matrix 
 \begin{equation}\label{disp-6}
M^{\a\d}=\chi^{\a\b\g\d}q_\b q_\g\,.
     \end{equation}
This matrix evidently satisfies the relations
\begin{equation}\label{disp-7}
M^{\a\d}q_\a=0\,,\qquad M^{\a\d}q_\d=0\,.
     \end{equation}
These relations have a clear physical meaning. The first equation represents the {\it charge conservation law}, while the second one means that an ansatz (\ref{disp-4}) with $q_\a\sim a_\a$ is a solution of (\ref{disp-5}). Certainly this solution is not physically meaningful, because it corresponds to a zero value of the field $\F_{\a\b}$, i.e., it is related to the {\it gauge invariance} of the field equations.    

Thus we are looking for a solutions of the system (\ref{disp-5}) constrained by the relations (\ref{disp-6}). On the matrix language, these relations mean that the columns and the rows of the matrix $M^{\a\d}$ are linearly dependent, i.e., the matrix is singular. Consequently, a system always has a non-zero solution. However we need more of that, in fact,  we are looking for an additional linear independent solution. Only this one will be of a physical meaning. 

It is an algebraic fact, that a linear system has two independent solution only if the adjoint of the characteristic matrix equal to zero. 
So we come to an equation 
\begin{equation}\label{disp-8}
(adj\, M)_{\a\b}=0\,.
\end{equation}
On a first view, it seems that we require here 16 conditions for 16 components of the matrix  $M^{\a\b}$. In fact, the situation is much different. It can be proved \cite{Itin:2009aa} that, for a matrix which satisfies the conditions   (\ref{disp-7}), the adjoint matrix is of the form 
 \begin{equation}\label{disp-9}
(adj\, M)_{\a\b}=\lambda(q)q_\a q_\b\,.
     \end{equation}
Consequently, a necessary condition for existence of a physically meaningful solution for a wave propagational system is expressed by a scalar equation 
\begin{equation}\label{disp-10}
\lambda(q)=0\,.
     \end{equation}
The function $\lambda(q)$ is a homogeneous 4-th order polynomial in the wave covector $q_\a$. It's explicit forms are given in 
\cite{birkbook}, \cite{Itin:2009aa}.

\section{An anisotropic dispersion relation}
For a generalized anisotropic media with a constitutive tensor (\ref{gen-9}), we apply an ordinary (1+3)-decomposition of the wave covector 
\begin{equation}\label{bian-1}
q^\a=(w,k^i)\,.
 \end{equation}
The characteristic matrix has now  the entries 
\begin{equation}\label{bian-2}
 M^{00}=\ep^{ij}k_ik_j\,, \quad M^{0i}=-\ep^{ij}k_jw\,, \quad M^{i0}=-\ep^{ij}k_jw\,,
     \end{equation}
and 
\begin{equation}\label{bian-3}
M^{ij}=-\ep^{ij}w^2+\epsilon^{imn}\epsilon^{jpq}\pi_{mp}k_nk_q\,.
  \end{equation}
We write the latter equation in a short form 
\begin{equation}\label{bian-4}
M^{ij}=\ep^{ij}w^2-\tau^{ij}\,.
\end{equation}
where a matrix $\tau^{ij}$ is defined as 
\begin{equation}\label{bian-5}
\tau^{ij}=\epsilon^{imn}\epsilon^{jpq}\pi_{mp}k_nk_q\,.
  \end{equation}
Due to the relations $\tau^{ij}k_i=\tau^{ij}k_j=0$ it is singular.

In correspondence with (\ref{disp-9}), it is enough to calculate only one component of the adjoint matrix. Write 
 \begin{equation}\label{bian-6}
(adj\, M)_{00}=\frac 1{3!}\epsilon_{i_1i_2i_3}\epsilon_{j_1j_2j_3}M^{i_1j_1}M^{i_2j_2}M^{i_3j_3}\,.
  \end{equation}
  Substituting (\ref{bian-4}) we derive for $(adj\, M)_{00}=\lambda w^2$
  \begin{eqnarray}\label{bian-7}
\lambda&=&\frac 1{3!}w^4 \left(\epsilon_{i_1i_2i_3}\epsilon_{j_1j_2j_3}\ep^{i_1j_1}\ep^{i_2j_2}\ep^{i_3j_3}\right)-\nonumber\\
&&\frac 1{2!}w^2\left(\epsilon_{i_1i_2i_3}\epsilon_{j_1j_2j_3}\ep^{i_1j_1}\ep^{i_2j_2}\tau^{i_3j_3}\right)+\nonumber\\
&&\frac 1{2!}\left(\epsilon_{i_1i_2i_3}\epsilon_{j_1j_2j_3}\ep^{i_1j_1}\tau^{i_2j_2}\tau^{i_3j_3}\right)\,.
\end{eqnarray}
In  the first term, we recognize  the determinant of the matrix $\ep$. In the second and the third terms, the adjoint of the matrices  $\ep$ and $\tau$ are extracted. 
  Consequently, the desired {\it dispersion relation} obtains a compact matrix form (the central dot denotes the matrix multiplication)
  \begin{equation}\label{bian-8}
w^4(det\,{\mathbf \varepsilon})-w^2 tr\left({\mathbf \tau}^T\cdot adj\,{\mathbf \varepsilon}\right)+tr\left({\mathbf \varepsilon^T}\cdot adj\,{\mathbf \tau}\right)=0
\end{equation}
This equation can be rewritten in the explicit tensorial form.  
The adjoint of the matrix $\tau$ is calculated by substituting (\ref{bian-5}) into the definition 
    \begin{equation}\label{bian-9}
    (adj\,\tau)_{ij}=\frac 1{2!}\epsilon_{ii_2i_3}\epsilon_{jj_2j_3}\tau^{i_2j_2}\tau^{i_3j_3}\,.
    \end{equation}
    The result is a scalar function multiplied by $k_ik_j$
       \begin{equation}\label{bian-10}
  (adj\,\tau)_{ij}=  \big((adj\, \pi)^{mn}k_mk_n\big)k_ik_j\,.
 \end{equation}    
For a invertible matrices  ${\mathbf \varepsilon}, {\mathbf \mu}$, we use (\ref {bian-10}) to  rewrite the expression (\ref{bian-8})  in a form
    \begin{equation}\label{bian-11}
 w^4-2\big(\psi^{ij}k_ik_j\big)w^2 +\frac{\ep^{ij}k_ik_j}{det\, \ep}\,\frac{\mu^{mn}k_mk_n}{det\, \mu}=0\,,
\end{equation}
where
\begin{equation}\label{bian-12}
\psi^{ij}=\frac 12\epsilon^{imn}\epsilon^{jpq}\ep^{-1}_{nq}\mu^{-1}_{mp}\,.
\end{equation}
Note some straightforward facts resulted from this expression:

{\bf (1)} The dispersion relation (\ref{bian-11}) is {\it symmetric} under interchange between $\ep$ and $\mu$. 

{\bf (2)} A criterion for the {\it absence of zero-frequency modes} with  $w=0$  for some $(k_1,k_2,k_3) \neq (0,0,0)$ takes the form: {\it The symmetric parts of the matrices $\ep$ and $\mu$ are definite (positive
 or negative).}\cite{vol}
 
 {\bf (3)} The necessary and sufficient condition for {\it hyperbolicity}  (four  real $w$ for any real $\bf{k}$) is expressed as a system of inequalities
 \begin{equation}\label{bian-13}
 \psi^{ij}k_ik_j>0\quad \rm{positive \, definite}
 \end{equation}
 and 
 \begin{equation}\label{bian-14}
 \big(\psi^{ij}k_ik_j\big)^2\ge \frac{\ep^{ij}k_ik_j}{det\, \ep}\,\frac{\mu^{mn}k_mk_n}{det\, \mu}>0\,.
 \end{equation}
 
{\bf (4)} In the {\it non-birefringence} case \cite{Itin:2005iv} with a unique optical metric, the dispersion relation (\ref{bian-11}) takes the form  
 \begin{equation}\label{bian-15}
 w^2-\psi^{ij}k_ik_j=0\,.
 \end{equation}
 Thus the optical metric is of Minkowski signature if and only if the matrix $\psi^{ij}$ is positive definite. 
\section{Examples}
\subsection{Isotropic case}
 It easy to check that in the isotropic case with 
 $ \ep_{ij}=\ep\d_{ij}\,, \mu_{ij}=\mu\d_{ij}$
   (\ref{bian-11}) yields the ordinary dispersion relation
 $(\ep\mu)w^2-k^2=0\,.$
  \subsection{Diagonal case}
  Let us consider a more involved example of two diagonal  matrices 
   \begin{equation}\label{ex-3}
  {\mathbf \ep}=diag(\ep_1,\ep_2,\ep_3)\,,\qquad  {\mathbf\mu}=diag(\mu_1,\mu_2,\mu_3)\,.
  \end{equation}
  The last term of (\ref{bian-11}) takes the form
  \begin{equation}\label{ex-4}
  \frac{\ep_1k_1^2+\ep_2k_2^2+\ep_3k_3^2}{\ep_1\ep_2\ep_3}\,\cdot\,\frac{\mu_1k_1^2+\mu_2k_2^2+\mu_3k_3^2}{\mu_1\mu_2\mu_3}\,.
   \end{equation}
  The coefficient of the second term of  (\ref{bian-11}) also easily calculated 
  \begin{equation}\label{ex-5}
  k_1^2\left(\!\frac 1{\ep_2\mu_3}\!+\!\frac 1{\ep_3\mu_2}\!\right)\!+ \!k_2^2\left(\!\frac 1{\ep_1\mu_3}\!+\!\frac 1{\ep_3\mu_1}\!\right)\!+\! k_3^2\left(\!\frac 1{\ep_1\mu_2}\!+\!\frac 1{\ep_2\mu_1}\!\right)
   \end{equation}
  Consequently, the dispersion relation in this case takes the form
   \begin{eqnarray}\label{ex-6}
 && \Big(\ep_1\ep_2\ep_3\mu_1\mu_2\mu_3\Big)w^4+\Big(\ep_1\mu_1k_1^2(\ep_2\mu_3+\ep_3\mu_2)+\nonumber\\
&&\ep_2\mu_2k_2^2(\ep_1\mu_3+\ep_3\mu_1)+\ep_3\mu_3k_3^2(\ep_1\mu_2+\ep_2\mu_1)\Big)w^2+\nonumber\\
&&
(\ep_1k_1^2+\ep_2k_2^2+\ep_3k_3^2)
(\mu_1k_1^2+\mu_2k_2^2+\mu_3k_3^2)=0
\end{eqnarray}
In a special case $\mu_i=1$, this formula coincides with the one given in \cite{LL},\cite{Jack}.  
   \subsection{Magnetized ferrite }
 
Magnetized ferrite materials are described by an isotropic dielectric constant 
\begin{equation}
{\mathbf \ep}=diag(\ep,\ep,\ep)\,.
\end{equation} 
Under influence of the magnetic field the initially isotropic magnetic matrix obtains an anisotropic modification. For a magnetic field directed as the $z$-axis \cite{Ishimaru}, 
\begin{equation}
{\mathbf{\mu}}=\left( \begin{array}{ccc}
\mu & iq & 0 \\
-iq & \mu& 0 \\
0 & 0 & \mu_0 \end{array} \right)\,.
 \end{equation}
The last term of (\ref{bian-11}) takes now the form 
\begin{equation}
\frac{k^2}{\ep^2}\,\cdot\,\frac {\mu(k_1^2+k_2^2)+\mu_0k_3^2}{\mu_0(\mu^2-q^2)}\,.
\end{equation}
The coefficient of the second term of (\ref{bian-11}) is easily calculated. 
The inverse matrix 
 \begin{equation}
 {\mathbf{\mu}}^{-1}=\frac 1{r\mu_0\mu}
\left( \begin{array}{ccc}
\mu & iq & 0 \\
-iq & \mu& 0 \\
0 & 0 & \mu_0 \end{array} \right) \quad {\rm{ where}}  \quad 
r=\frac {\mu^2-q^2}{\mu\mu_0}\,. 
 \end{equation}
 Consequently the second term of (\ref{bian-11}) takes the form 
\begin{eqnarray}
&&(\d^{mp}\d^{nq}-\d^{mq}\d^{np})\frac {w^2}\ep \mu^{-1}_{mp}k_nk_q=
\nonumber\\
&&
\qquad \frac {w^2}{\mu_0\ep r}\Big[\mu(k_1^2+k_2^2)+\mu_0k_3^2\Big]\,.
\end{eqnarray}
The resulting dispersion relation is 
 \begin{eqnarray}
&& r\mu^2_0\ep^2 w^4-w^2\ep\mu_0\big[2k^2+(r-1)(k^2-k_3^2)\big]+
\nonumber\\
&&\qquad 
 k^2\left[k^2+\left(\frac{\mu_0}\mu-1\right)k_3^2\right]=0\,.
 \end{eqnarray}
In absent of an exterior magnetic field, $q=0$ and $\mu=\mu_0$. Consequently, $r=1$ and the isotropic dispersion relation reinstated. 

\section{Conclusion} 
A short dispersion relation is derived for a generalized anisotropic media. The corresponded matrices are not required  to be  symmetric, positive definite, and even invertible. This compact form is much simpler for calculation and for theoretical analysis than the ones represented in the electromagnetic literature \cite{IEEE1}, \cite{IEEE2}. 
The algebraic consideration presented here can be also useful for a compact algebraic representation of a general bianisotropic tensorial  dispersion relation represented recently in \cite{Obukhov:2004zz}. 
\begin{acknowledgments}
I thank F.-W. Hehl, V. Perlick, Y. Obukhov and Y. Friedman for most useful discussions. 
\end{acknowledgments}


\end{document}